\documentclass[10pt,prl,aps,twocolumn,showpacs,superscriptaddress]{revtex4}
\usepackage{graphicx}

\begin{document}

\title{Ground state phases of the Half-Filled One-Dimensional Extended 
Hubbard Model}

\author{Anders W. Sandvik} 
\affiliation{Department of Physics, Boston University, 
590 Commonwealth Avenue, Boston, Massachusetts 02215}
\affiliation{Department of Physics, {\AA}bo Akademi University, 
Porthansgatan 3, FIN-20500 Turku, Finland}
\affiliation{Department of Physics, University of California,
Santa Barbara, California 93106}

\author{Leon Balents} 
\affiliation{Department of Physics, University of California,
Santa Barbara, California 93106}

\author{David K. Campbell}
\affiliation{Department of Physics, Boston University, 
590 Commonwealth Avenue, Boston, Massachusetts 02215}
\affiliation{Department of Electrical and Computer Engineering, 
Boston University,\\ 44 Cummington Street, Boston, Massachusetts 02215}

\date{\today}


\begin{abstract}
Using quantum Monte Carlo simulations, results of a strong-coupling expansion,
and Luttinger liquid theory, we determine quantitatively the ground state 
phase diagram of the one-dimensional extended Hubbard model with on-site and 
nearest-neighbor repulsions $U$ and $V$. We show that spin frustration 
stabilizes a bond-ordered (dimerized) state for $U\approx V/2$ up to 
$U/t \approx 9$, where $t$ is the nearest-neighbor hopping. The transition from
the dimerized state to the staggered charge-density-wave state for large 
$V/U$ is continuous for $U \alt 5.5$ and first-order for higher $U$.
\end{abstract}

\maketitle

The one-dimensional Hubbard model, which describes electrons on a
tight-binding chain with single-particle hopping matrix element $t$ and 
on-site repulsion $U$, has a charge-excitation gap for any $U > 0$ at 
half-filling \cite{liebwu}. In the spin sector, the low-energy spectrum 
maps onto that of the $S=1/2$ Heisenberg chain; the spin coupling 
$J=4t^2/U$ for $U \to \infty$. The spin spectrum is therefore gapless 
and the spin-spin correlations decay with distance $r$ as $(-1)^r/r$ 
\cite{luther}. Hence, the ground state is a quantum critical staggered 
spin-density-wave (SDW). In the simplest {\it extended Hubbard model},
a nearest-neighbor repulsion $V$ is also included. The Hamiltonian is,
in standard notation and with $t=1$ hereafter,
\begin{eqnarray}
H & = & -t\sum_{\sigma=\uparrow,\downarrow}\sum_{i}
(c^{\dagger}_{\sigma,i+1}c_{\sigma,i} +
c^{\dagger}_{\sigma,i}c_{\sigma,i+1}) \nonumber\\ & + & U\sum_i
n_{\uparrow,i}n_{\downarrow,i} + V\sum_i n_in_{i+1}.
\label{ham}
\end{eqnarray}
The low-energy properties for $V \alt U/2$ are similar to those at
$V=0$.  For higher $V$ the ground state is a staggered charge-density-wave 
(CDW), where both the charge and spin excitations are gapped. The
transition between the critical SDW and the long-range-ordered CDW has
been the subject of numerous studies
\cite{bari,emery1,solyom,hirsch,cannon,vandongen,voit,nakamura,sengupta,sandvik1,tsuchiizu,jeckelmann}. Until recently, it was believed that the SDW-CDW 
transition occurs for all $U>0$ at $V \agt U/2$ and that it is continuous for 
small $U$ ($\alt 5$) and first-order for larger $U$. However, based on a
study of excitation spectra of small chains, Nakamura argued that there is 
also a bond-order-wave
(BOW) phase \cite{nakamura}, where the ground state has a
staggered modulation of the kinetic energy density (dimerization), in
a narrow region between the SDW and CDW phases for $U$ smaller than
the value at which the transition changes to first order. Previous
studies \cite{hirsch,cannon,vandongen,voit} had indicated an SDW state
in this region. Nakamura's BOW-CDW boundary coincides with the
previously determined SDW-CDW boundary. The presence of dimerization
and the accompanying spin gap were subsequently confirmed using quantum
Monte Carlo (QMC) simulations \cite{sengupta,sandvik1}. The BOW phase now 
also has a weak-coupling theory \cite{tsuchiizu}.

The existence of an extended BOW phase has recently been disputed. Jeckelmann
argued, on the basis of density-matrix-renormalization-group (DMRG)
calculations, that the BOW exists only on a short segment of the 
first-order part of the SDW-CDW boundary \cite{jeckelmann}, 
i.e., that the transition always is 
SDW-CDW and that BOW order is only induced on part of the coexistence curve. 
However, QMC calculations demonstrate the existence of BOW order well away
from the phase boundary \cite{sandvik1}. 

Although several studies agree on the {\it existence} of an extended
BOW phase \cite{nakamura,sengupta,tsuchiizu,sandvik1}, the shape of
this phase in the $(U,V)$ plane has not yet been reliably
determined. The system sizes used in the exact diagonalization study
\cite{nakamura} were too small for converging the SDW-BOW boundary
(i.e., the spin gap transition) for $U \agt 4$. In the previous QMC
studies \cite{sengupta,sandvik1}, the emphasis was on verifying the
presence of BOW order and the phase transitions for $U \approx
4$. In this Letter, we present the complete phase diagram. Taking
advantage of recent QMC algorithm developments---stochastic series
expansion with directed-loop updates \cite{directed} in
combination with the quantum generalization \cite{sengupta} of the
parallel tempering method \cite{tempering}---we have carried out
high-precision, large-chain (up to $L=1024$) calculations for
sufficiently high $U$ ($\le 12$) to locate the point at which the BOW
order vanishes. In agreement with Ref.~\cite{jeckelmann}, we find
that the BOW exists also above the $U$ at which the transition to the
CDW state becomes first-order.  However, long-range BOW order exists
also below this point, and hence {\it the point at which the nature of
the transition changes from continuous to first-order is on the 
BOW-CDW boundary}. 

The phase diagram we find here is qualitatively similar to that obtained in 
a 4th order strong-coupling expansion, where the transition to the CDW state 
is determined by comparing the energies of the large-$V$ CDW 
state and the effective spin 
model including nearest- and next-nearest-neighbor interactions $J$ and $J'$
\cite{vandongen}. The BOW phase corresponds to the spontaneously dimerized 
phase of the spin chain, i.e., $J'/J > 0.241$ \cite{okamoto}. In Fig.~1, we 
compare our QMC phase boundaries with the strong-coupling result; the 
procedures giving the QMC boundaries will be discussed below. We will 
show that the system is a Luther-Emery liquid on the continuous BOW-CDW 
boundary, i.e., the charge gap vanishes and the spin gap remains
open. Evidence supporting this type of transition was also presented
in Ref.~\cite{sengupta}. Here we will further argue that the change to 
a first-order transition corresponds to the Luttinger charge exponent 
$K_\rho$ reaching the value $1/4$.

\begin{figure}
\includegraphics[width=6cm]{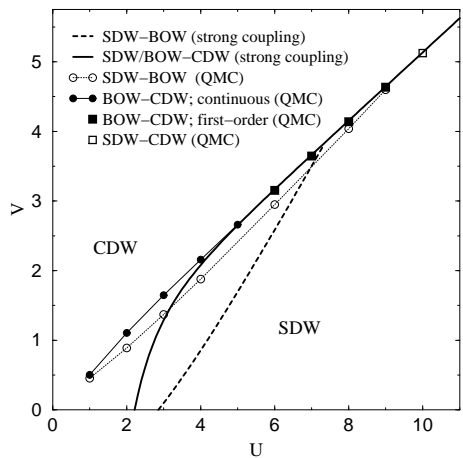}
\caption{QMC and strong-coupling phase diagram. The BOW is located 
between the SDW-BOW and BOW-CDW curves.}
\label{vc}
\end{figure}

\begin{figure}
\includegraphics[width=7cm]{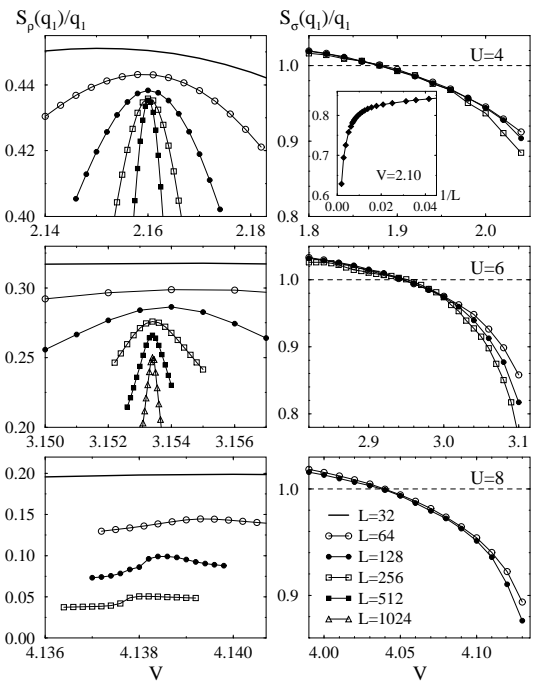}
\caption{Long-wavelength charge (left panels) and spin (right panels)
structure factors vs $V$ for $U=4$ (top), $6$ (middle), and $8$ (bottom).
The system sizes are indicated in the low-right panel. The $U=4$ inset 
shows the dependence on the inverse lattice size at $V=2.10$.}
\label{sq1}
\end{figure}

We extract the SDW-BOW and BOW-CDW boundaries using the
charge and spin exponents $K_\rho$ and $K_\sigma$. If there
is a spin or charge gap, the corresponding exponent vanishes. Otherwise
the equal-time correlation function $C_{\rho}(r) \sim  r^{-(K_\sigma+K_\rho)}$,
$C_{\sigma}(r) \sim r^{-(K_\sigma^{-1}+K_\rho)}$. If non-zero, the spin 
exponent $K_\sigma=1$ as a consequence of spin-rotation invariance 
\cite{voit2}. On periodic chains the exponents can be conveniently extracted 
from the static structure factors $S_{\rho,\sigma}(q)$
\cite{clay},
\begin{equation}
S_{\rho,\sigma} (q)={1\over L}\sum\limits_{j,k}{\rm e}^{iq(j-k))}
\langle (n_{\uparrow j} \pm n_{\downarrow j}) (n_{\uparrow k} \pm
n_{\downarrow k})\rangle,
\end{equation}
in the limit $q\to 0^+$:
\begin{equation}
K_{\rho,\sigma} = S_{\rho,\sigma}(q_1)/q_1,~~q_1=2\pi/L,~~L \to \infty .
\label{krhofroms}
\end{equation}
If there indeed are three successive phases, 
SDW-BOW-CDW, as $V$ is increased at a fixed value of $U$, then
the spin exponent $K_\sigma =1$ in the SDW phase and $K_\sigma =0$ everywhere
else. The charge exponent $K_\rho =0$ everywhere, except exactly at the 
BOW-CDW transition point if this is a continuous quantum phase transition 
(i.e., if the charge gap vanishes). In contrast, if the transition is 
first-order, then $K_\rho =0$ also on the phase boundary. Using the
relation (\ref{krhofroms}) for a finite system, any
discontinuities will naturally be smoothed, and one can only
expect to observe $S_{\rho,\sigma}(q_1)/q_1$ developing sharp
features as $L$ is increased. In Fig.~\ref{sq1} we show results
demonstrating this for several different system sizes at $U=4$, $6$,
and $8$.

Looking first at the charge exponent, if $K_\rho > 0$ on the
BOW-CDW boundary and $K_\rho = 0$ elsewhere, then one can expect a peak
developing in $S_{\rho}(q_1)/q_1$ versus $V$. The peak position
corresponds to the critical $V$, and the peak height should converge
to $K_\rho$. If the transition is first-order, $S_{\rho}(q_1)/q_1$
should converge to zero for all $V$, but one can still expect some
structure at the phase boundary for finite $L$ as the nature of the
ground state changes. In Fig.~\ref{sq1}, for $U=4$ and $6$ the
development of sharp peaks is apparent. For $U=4$ the peak-height
converges to a non-zero value, implying a continuous transition at $V
\approx 2.160$ with $K_\rho \approx 0.43$. For $U=6$ the convergence
to a value $>0$ is not clear, but the transition point is given
accurately by the peak-position, which shows very little
size-dependence.  It has been shown previously that the transition is
first-order for $U=6$ \cite{hirsch,sengupta}, and the peak should therefore
in fact converge to zero. The rather slow decay reflects the proximity 
to the point at which the transition becomes continuous. For $U=8$ the 
peak does not sharpen, but instead a step develops at the critical $V$. 
The whole curve converges to $0$ as $L \to \infty$. The transition is hence 
strongly first-order in this case, in agreement with previous calculations. 
As seen in Fig.~\ref{vc}, and as observed already by Hirsch \cite{hirsch}, 
the locations of the $U=6$ and $8$ critical points agree very well with the 
strong-coupling expansion \cite{note1}.

In the SDW phase, one cannot expect to
easily find $S_{\sigma}(q_1)/q_1 \to 1$ exactly,
due to logarithmic corrections that affect various quantities strongly 
even for very long chains \cite{woynarovich,eggert1}. However, the 
log-corrections are known to vanish in the frustrated $J-J'$ spin chain at 
its dimerization transition \cite{eggert2}, and hence, since the SDW-BOW 
transition should be of the same nature, the log-corrections should vanish 
here as well. The transition at fixed $U$ should therefore be signaled
by $S_{\sigma}(q_1)/q_1$ crossing $1$ from above as $V$ is
increased. Because of the vanishing log-corrections at the transition,
the crossing point with $1$ should not move significantly as $L$ is
increased, but the drop below $1$ should become increasingly sharp, and
eventually $S_{\sigma}(q_1)/q_1$ should approach $0$ inside the BOW
phase. This method was used in Ref.~\cite{sengupta} and gave a
slightly higher critical $V$ for the SDW-BOW transition at $U=4$ than
the exact diagonalization \cite{nakamura}. We now have results for a
wider range of couplings. The results shown in Fig.~\ref{sq1} are in
accord with the above discussion for all three $U$-values;
$S_{\sigma}(q_1)/q_1$ crosses $1$ at a $V$-point which does not move
visibly between $L=64$ and $L=256$. For larger $V$, one can see a 
sharper drop for the larger system sizes. The size dependence at $U=4,V=2.1$ 
is shown in an inset. Here the convergence to $0$, i.e., 
the presence of a spin gap, is apparent. If the spin gap is small, as it 
is close to the phase boundary, the convergence to $0$ will obviously 
occur only for very large systems.

Results such as those shown in Fig.~\ref{sq1} were used to determine
the phase boundaries in Fig.~\ref{vc}. As already noted, the BOW aspect 
of the phase diagram differs from previous proposals
\cite{nakamura,sengupta,tsuchiizu,jeckelmann} in that the BOW-CDW transition
can be either continuous or first-order, i.e., {\it the change of order
occurs on the BOW-CDW boundary}. The existence of two special points,
one where the transition-order changes and one at higher $U$ where the
BOW vanishes, was also suggested by Jeckelmann \cite{jeckelmann}, who,
however, insists that the BOW does not exist for
small $U$ where the transition to the CDW state is continuous (i.e.,
his phase diagram has no continuous BOW-CDW transition). 
We have carried out calculations for $U$ down to
$1$, and, as shown in Fig.~\ref{vc}, we still find a BOW phase
there. Most likely, in view also of weak-coupling arguments
\cite{tsuchiizu}, the BOW extends down to $U=0^+$. We find no BOW for
$U \agt 9$. In the strong-coupling expansion, using the
couplings $J$ and $J'$ derived by van Dongen \cite{vandongen}, the
effective spin model is gapped, i.e., $J' > 0.241 J$ \cite{okamoto},
above the dashed curve in Fig.~\ref{vc}.  The $J-J'$ mapping is not
applicable beyond the transition into the CDW state, which
(the solid curve in Fig.~\ref{vc}) was previously calculated by 
comparing the 4th-order CDW and $J-J'$ energies \cite{vandongen}. 
The 4th-order BOW region ends at $U
\approx 7$, where the spin-gap curve crosses the CDW-transition
curve. This is slightly lower than what we find based on QMC. The
strong-coupling BOW extends down to smaller $U,V$, but clearly the
4th-order result cannot be expected to be quantitatively accurate in
this region. Nevertheless, the spin-frustration mechanism
consistently explains the presence of an SDW-BOW transition and an
extended BOW phase. Spin-frustration was previously cited by Jeckelmann 
\cite{jeckelmann}, but, surprisingly, he used it in support of a BOW 
of vanishing extent.

\begin{figure}
\includegraphics[width=6cm]{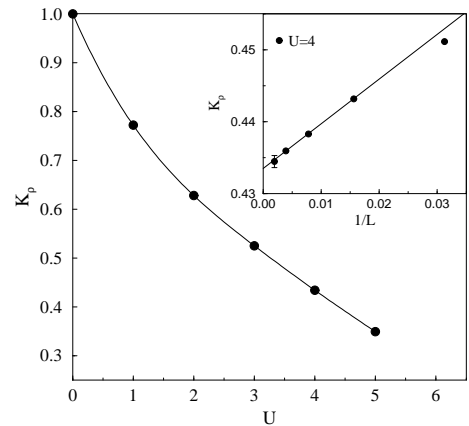}
\caption{QMC results for Luttinger charge exponent on the BOW-CDW boundary
(solid circles). The inset shows the finite-size scaling for $U=4$.}
\label{kc}
\end{figure}

Next, we consider the nature of the BOW-CDW transition.
As discussed in Refs.~\cite{nakamura,tsuchiizu}, the continuous critical 
point for small U is described by a Gaussian free (charge) boson theory,
characterized by the parameter $K_\rho$.  At generic values of 
(repulsive) $U,V$, the leading ``$4k_F$'' umklapp process is present, 
and has scaling dimension $\Delta_{4k_F}=2K_\rho$ and is hence relevant 
($\Delta_{4k_F}<2$) for $K_\rho<1$.  At the BOW-CDW transition, this operator
vanishes, leading to a vanishing of the charge gap.  For consistency, no other
relevant operators should be present, which would otherwise require fine 
tuning to zero, making the Gaussian theory a multicritical point. The most 
dangerous candidate is the ``$8k_F$'' umklapp process, with $\Delta_{8k_F}=
4\Delta_{4k_F}=8K_\rho$, so a continuous Gaussian critical point is possible 
only for $1>K_\rho>1/4$.  Even in this range, the Gaussian theory is an 
unusual critical point with {\it non-universal} behavior, e.g., the 
correlation length exponent $\nu=1/(2-2K_\rho)$. 

Extrapolated QMC results 
for $K_\rho$ on the BOW-CDW boundary are shown in Fig.~\ref{kc}. 
The finite-size corrections appear to be of the form $1/L^\alpha$, with
an $U$ dependent exponent $\alpha$. At $U=4$, $\alpha \approx 1$, as
shown in the inset of Fig.~\ref{kc}. For larger $U$, $\alpha$ decreases
rapidly and is difficult to determine accurately for $U \agt 5$. The extrapolated
$K_\rho$ value at $U=5$  in Fig.~\ref{kc} should be regarded as an upper bound. 
At $U=6$, the extrapolated $K_\rho < 1/4$, and hence we expect an eventual 
drop to $0$. This is consistent with clear signals of a first-order transition
\cite{sengupta}. Also at $U=5.5$ there are signs of first-order behavior, 
e.g., in order parameter histograms such as those considered in 
Ref.~\onlinecite{sengupta}. We believe that the change from a continuous 
to a first-order transition occurs between $U=5$ and $5.5$.

What is the nature of the tricritical point at which the transition
becomes first-order?  The simplest scenario is that this is the last 
(marginally) stable point of the Gaussian fixed line, i.e. with $K_\rho=1/4$.
This hypothesis predicts that the critical $K_\rho$ continuously approaches 
$1/4$ as the tricritical point $U=U_{t}$ is approached from below, as 
$K_\rho -1/4 \sim \sqrt{(U_{t}-U)/U_{t}}$. We do not have sufficient 
data to verify this form, but it is consistent with a sharp drop to $0$ 
between $U=5$ and $5.5$ (Fig.~\ref{kc}), required since at $U=5.5$ the transition should 
be first-order. Hence, we favor this behavior over the a priori consistent 
(but less simple) possibility of a non-trivial ``strong coupling'' tricritical
fixed point far from the Gaussian line. 

\begin{figure}
\includegraphics[width=7.25cm]{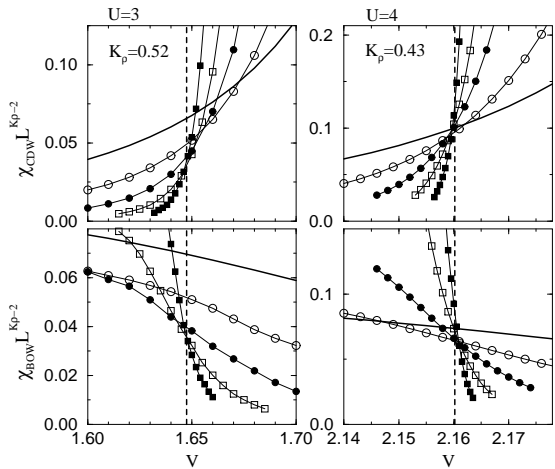}
\caption{Finite-size scaling of the BOW and CDW susceptibilities for
$U=3$ (left) and $4$ (right). The symbols correspond to different system 
sizes as in Fig.~\ref{sq1}. The dashed lines indicate the independently 
determined critical points.}
\label{cross}
\end{figure}

To further demonstrate the Luther-Emery state on the continuous BOW-CDW 
curve, we study the finite-size scaling of the CDW and BOW susceptibilities, 
$\chi_{\rm CDW}(\pi)$ and $\chi_{\rm BOW}(\pi)$ (with their standard 
Kubo-integral definitions \cite{sengupta}). Both the charge and bond 
correlations should decay as $(-1)^rr^{-K_\rho}$ \cite{voit2}, implying that 
the susceptibilities scale with system size as $L^{2-K_\rho}$. Thus,
$\chi(\pi) L^{K_\rho-2}$ curves for different $L$ should intersect at the 
critical BOW-CDW point. Fig.~\ref{cross} shows results for $U=3$ and $4$, 
using the $K_\rho$ values determined above. For $U=4$ the expected scaling 
can be observed even for small systems. For $U=3$ the corrections are larger,
and the asymptotic scaling sets in only for $L \agt 128$. This is clearly due
to the smaller spin gap at $U=3$, which implies a longer length-scale below which 
remaining spin correlations affect the charge and bond fluctuations.

In summary, we have determined the ground state phase diagram of the
extended Hubbard model at half-filling. The dimerized BOW phase can be
explained by spin-frustration. The BOW-CDW transition changes from 
continuous to first-order between $U=5$ and $5.5$. On the critical $(U,V)$ 
curve the system is a Luther-Emery liquid, with a charge exponent $K_\rho$ 
decreasing from $1$ as $U$ is increased from $0$. We have argued that the 
minimum $K_\rho =1/4$ and that the BOW-CDW transition becomes first-order 
when this value is reached.

This work was supported by the Academy of Finland, project No.~26175
(AWS), by the NSF under Grants No.~DMR-9985255 (LB) and DMR-97-12765 (DKC),
and by the Sloan and Packard foundations (LB). The numerical calculations 
were carried out at the CSC in Espoo, Finland, and at the NCSA in Urbana, 
Illinois.

\null


\begin{thebibliography}{00}
\bibitem{note1} The $U=8$, $L=256$ data are not completely reproducible: 
The location of the $S_{\rho}(q_1)/q_1$ maximum fluctuates slightly between
runs, due to meta-stability at the strongly first-order transition. The 
parallel tempering scheme \cite{sengupta} overcomes this problems completely
only for smaller $L$ and/or smaller $U$. For $U=4$ and $6$, the results are 
reproducible for all $L$ considered.
\end{thebibliography}
\end{document}